\documentclass[
aps,pra,
reprint,
a4paper,
superscriptaddress,
longbibliography,
preprintnumbers,
]{revtex4-1}
\usepackage[utf8]{inputenc}
\usepackage[T1]{fontenc}
\usepackage{amssymb,amsmath,amsfonts,amsthm}
\usepackage{braket}
\usepackage{hyperref}
\usepackage{epsfig}
\hypersetup{citecolor=magenta}
\hypersetup{colorlinks=true}
\hypersetup{linkcolor=blue}
\hypersetup{urlcolor=blue}

\def\<{\langle}\def\>{\rangle}

\newcommand\C{{\mathbb C}}
\newcommand\B{{\mathsf B}}
\newcommand\Act{{\mathsf{Act}}}

\newcommand\map[1]{{\mathcal #1}}
\newcommand\Chi{{\mathrm X}}
\newcommand\Part[1]{{\mathbb #1}}
\newcommand\Supp{{\mathsf{Supp}}}
\DeclareMathOperator\Tr{tr}

\newtheorem{ax}{{Axiom}}
\newtheorem{theo}{{Theorem}}
\newtheorem{lem}{{Lemma}}
\newtheorem{cor}{{Corollary}}
\newtheorem{defi}{{Definition}}

\newcounter{biscompt}
\newtheorem{theobis}[biscompt]{Theorem}

\begin{document}

\author{Giulio Chiribella}
\email{giulio@cs.hku.hk}
\affiliation{QICI Quantum Information and Computation Initiative, Department of Computer Science, The University of Hong Kong, 
Pokfulam Road, Hong Kong}
\affiliation{Department of Computer Science, Parks Road, Oxford, OX1 3QD, UK}
\affiliation{Perimeter Institute for Theoretical Physics, Waterloo, Ontario N2L 2Y5, Canada}

\author{Ad\'an Cabello}
\email{adan@us.es}
\affiliation{Departamento de F\'{\i}sica Aplicada II,
Universidad de Sevilla, E-41012 Sevilla, Spain}
\affiliation{Instituto Carlos~I de F\'{\i}sica Te\'orica y Computacional, Universidad de Sevilla, E-41012 Sevilla, Spain}

\author{Matthias Kleinmann}
\email{matthias.kleinmann@uni-siegen.de}
\affiliation{Department of Theoretical Physics,
University of the Basque Country UPV/EHU,
P.O.\ Box 644, E-48080 Bilbao, Spain}
\affiliation{Naturwissenschaftlich--Technische Fakult\"at,
Universit\"at Siegen,
Walter-Flex-Stra{\ss}e 3, 57068 Siegen, Germany}

\author{Markus P.\ M\"uller}
\email{markus.mueller@oeaw.ac.at}
\affiliation{Institute for Quantum Optics and Quantum Information, Austrian Academy of Sciences, Boltzmanngasse 3, A-1090 Vienna, Austria}
\affiliation{Perimeter Institute for Theoretical Physics, Waterloo, Ontario N2L 2Y5, Canada}


\title{General Bayesian Theories and the Emergence of the Exclusivity Principle}


\begin{abstract}
We address the problem of reconstructing  quantum theory from the perspective of an agent who makes bets about the outcomes of possible experiments. 
We build  a general Bayesian framework that can be used  to organize the agent's beliefs and update 
them when new information becomes available. Our framework includes as special cases classical and quantum probability theory, as well as 
 other forms of probabilistic reasoning that may arise in future physical theories.  
Building on this framework, we  develop 
a notion of ideal experiment, which in   quantum theory coincides with   the   notion of    projective measurement. 
We then prove that, in every general Bayesian theory,   ideal experiments must satisfy the exclusivity principle, a property of projective measurements that plays a central role in the characterization of   quantum correlations.  Our result suggests that the set of quantum correlations may be completely characterized in terms of  Bayesian consistency conditions. 
\end{abstract}


\maketitle


{\em Introduction.---}%
Quantum theory portrays a world where the outcomes of individual measurements cannot be predicted with certainty. And yet, the quantum predictions are strikingly accurate and explain an astonishingly broad range of phenomena. The reason for this broad applicability still remains controversial. Does the quantum formalism describe how Nature works at the fundamental level?  
Or it is just a mathematical tool for guessing  the outcomes of our experiments? 

Albeit with a variety of nuances, different interpretations of quantum theory 
tend to favor either one or the other view. For example, Everett's 
interpretation \cite{Everett57} holds that the quantum framework refers to a 
multitude of universes interfering with each other. On the other hand of the spectrum, 
QBism, which originally  stood for quantum Bayesianism \cite{FS13}, views quantum theory as a set of 
rules that constrain how agents should make  bets about the outcomes of their
experiments.

Different interpretations are reflected into different ways to 
understand  quantum correlations. Since  Bell \cite{bell1964einstein}, it has been known   that quantum correlations are incompatible with the intuitive worldview known as local realism. But intriguingly, the quantum violations of Bell's inequalities are not maximal: more general theories compatible with relativistic causality could in principle lead to larger violations \cite{khalfin1992quantum,popescu1994quantum,rastall1985locality}. Following up on this observation, various physical principles have been proposed to explain the quantum bounds on correlations \cite{brassard2006limit,linden2007noquantum,pawlowski2009information,navascues2010glance,fritz2013local}. Behind this approach lies the idea that the quantum bounds should be explained in terms of principles constraining how Nature behaves.    However, this is not the only option. Instead, one could search for principles constraining how agents assign probabilities to the outcomes of their experiments.
 This approach has remained mostly
unexplored so far, partly due to the lack of a suitable
framework, and partly due to a widespread belief that quantum correlations require new physical principles. 
 Even within QBism,
the Born rule is regarded as ``an empirical
{\em addition} to the laws of Bayesian probability'' \cite{FS13}, rather
than a consequence of Bayesian probability itself.

In this Rapid Communication we demonstrate that  a surprisingly large portion of the set of quantum correlations follows directly from elementary Bayesian conditions.   We first  build a general Bayesian framework, describing the  ways in which an agent  can  update its beliefs when a new piece of information becomes available. The framework assumes only basic Bayesian laws, such as the validity of the rule of conditional probability, and the consistency of beliefs at different moments of time. 
  Surprisingly, we find  that these rather minimalistic assumptions imply the validity of the   { exclusivity principle}, a  feature of quantum theory that characterizes a large portion of the set of  quantum correlations \cite{CSW10, Cabello13, Yan13,ATC14, 
Cabello15, Henson15, AFLS15,Cabello18}.


The exact  statement of the exclusivity principle will be given later in the paper. For the moment, the crucial observation is that  the exclusivity principle  {\em does not hold for arbitrary  experiments}.  In quantum theory, it holds for projective measurements, but it fails to hold for certain nonprojective measurements  \cite{AFLS15,CY16}.  This means that, in order to formulate the exclusivity principle in a general physical theory, one has first to extend the notion of projective measurement beyond the quantum framework. 
Such extension is far from straightforward, because  there are multiple inequivalent notions that  generalize  of the notion of projective measurement in quantum theory   \cite{CY16}. On top of that, once a choice is made, the exclusivity principle may or may not hold, depending on the theory under consideration.  For example, Ref.   \cite{CSW10}  showed that, for a certain generalization of the notion of projective measurement, the exclusivity principle is violated by theories that predict superstrong correlations such as Popescu-Rohrlich boxes \cite{popescu1994quantum}. 

Theories that satisfy the exclusivity principle have a remarkable property: under  the natural assumption that two statistically independent experiments can be performed in parallel,  the correlations arising in every Bell and Kochen-Specker contextuality scenario  must be contained in the quantum set \cite{Cabello18}. This means that, under a  mild assumption, the  quantum set for these contextualty scenarios can be characterized as the largest set of correlations compatible with the exclusivity principle.

In this paper we derive the exclusivity principle from Bayesian consistency conditions. We first introduce a general framework for Bayesian theories. 
We then formulate a general notion of an  ideal experiment, which in the special case of quantum theory coincides with the notion of  projective measurement. 
Our first result is that ideal experiments exist in  every general Bayesian theory (Theorem 1).  Our second result is that the outcome probabilities arising from  ideal experiments satisfy  the exclusivity principle (Theorem 2). 
Complemented with  results of Ref. \cite{Cabello18}, our results imply that the quantum set of correlations for any Bell or Kochen-Specker contextuality scenario can be derived just from elementary consistency conditions on the agent's probability assignment. This conclusion surpasses the expectations of QBism:  rather than being an empirical addition  \cite{FS13}, the Born rule appears as a consequence  of Bayesian consistency conditions alone.


{\em Beliefs and probabilities.---}%
Consider the situation of an agent who makes bets about the outcomes of  experiments 
on a given physical system.  We start from a basic class of experiments, which we call {\em principal experiments}. 
The outcomes of the principal experiments 
form a single sample space $\Chi$, equipped with a $\sigma$-algebra of events $\Sigma$, namely a collection of subsets of $\Chi$ satisfying the properties (i) $\Chi \in \Sigma$, (ii) $E \in \Sigma$ implies $(\Chi\setminus E)\in \Sigma$, and (iii) $(E_i)_i\subset \Sigma$ implies $\bigcup_i E_i \in \Sigma$. In typical cases, $\Chi$ is a finite set and $\Sigma$ is the power set of $\Chi$. 
A {principal experiment} corresponds to a partition $\Part E = (E_i)_i$ of the sample space $\Chi$ 
into disjoint events. For brevity, we will identify  the experiment and the 
corresponding partition. 

In making a bet, the agent will rely on its beliefs, including beliefs on the 
laws of physics, or beliefs on the prior history of the physical system 
involved in the bet. We denote by $\B$ the set of all possible beliefs. For a 
given belief $\beta$,  
the agent  assigns a probability distribution $p:  E \mapsto p(E|\beta)$ satisfying the usual conditions (i) $p(E|\beta) \ge 0$ 
for all events $E$, (ii) $p(\bigcup_i E_i|\beta)= \sum_i p (E_i|\beta)$ 
whenever all $E_i$ are mutually disjoint, and (iii) $p(\Chi|\beta)= 1$.   
Here we make the standard assumption that the probability of an event $E$ is independent of  the specific experiment $\Part E$ in  which $E$ arises. 

We stress that the belief $\beta$ determines the probability assignment $p(E|\beta)$, but not {\em vice versa}: in general, a belief contains much more than just the outcome probabilities of principal experiments.  For example,  we will see that in quantum theory  a belief  is described by a density matrix, while the  principal experiments are commuting projective measurements, corresponding to projectors that are diagonal with respect of a fixed standard basis.  
The probability assignment  $p(E|\beta)$ depends only on the diagonal entries of the density matrix $\beta$, 
  and therefore it is not sufficient to determine the whole matrix. 


{\em Bayesian updates.---}%
Suppose that the agent 
receives a guarantee that an event $E$ is the case. 
As a consequence, the agent will update its old belief $\beta$ into a new belief, which we denote  as $\beta' = E \beta$. 
 Again, we make the standard assumption that the new belief is independent of the specific experiment in which $E$ arises. 


The point of updating beliefs is to compute conditional probability distributions. We demand that the probability assignment for the updated belief 
$E\beta$ is given by the rule of conditional probabilities:
\begin{ax}[Rule of conditional probabilities]\label{ax:condprob}
For every initial belief $\beta$, and for every pair of events $E$ and $F$ with $p(E|\beta)\not  = 0$, the updated belief $E\beta$ satisfies the rule of conditional probabilities
\begin{equation}\label{bayes}
 p (F|E \beta) = \frac{p (E \cap F|\beta) }{p (E|\beta)} \, .
\end{equation}
\end{ax}
The rule of conditional probabilities implies several properties of the update map $\beta \mapsto E \beta$. For example, it implies that, once the agent updates its belief based on the event $E$, the agent becomes certain of the event $E$. Indeed, one has $p(E| E \beta) = 1$, which follows from letting $E=F$ in Eq.~\eqref{bayes}. 

We stress that the update map does not represent a physical process on the observed system, but rather an operation internal to the agent. In the following, we formulate two conditions that the update should satisfy in order for the agent to be consistent with its beliefs. 


{\em Forward consistency.---}%
Suppose that the agent is certain of the event $E$, namely $p(E|\beta)=1$.   In this case, a guarantee that $E$ is the case
does not add any new information, and  should not lead to any update:
\begin{ax}[Forward consistency]\label{ax:forward} If the agent is certain of the event $E$, 
then 
 the agent's belief does not change under the update for event  $E$. Mathematically: for every $\beta \in \B$ and every $E\in \Sigma$, $p(E|\beta)=1$ implies $E\beta= \beta$. 
\end{ax}
%
Forward consistency constrains the agent in how it updates the belief forward in time. 
A simple consequence of this constraint is that the total event $E=\Chi$ does not lead to any update, namely  $\Chi\beta= \beta, \, \forall \beta\in \B$.


{\em Actions.---}%
So far, we considered the situation where an agent bets directly on the occurrence of a certain event. More generally, the conditions under which a bet is made can be altered by some action. 
We use ``action'' broadly, including situations in which the agent lets the system evolve 
under its natural dynamics. 

We denote the set of all actions as $\Act$, and we assume that it is  a monoid,  meaning that {(i)} actions can be  composed with one another, {(ii)} the composition is associative, and {(iii)} there exists an identity action  \cite{kraemer2018operational, 
chiribella2018agents}.   When an action $\map A$ is performed, the agent  generally changes its belief to a new belief $\beta' = \map A \beta$.  We assume that the belief change satisfies the conditions (i) $ (\map A \map B)  \beta  =   \map A  \, (\map B \beta), \forall \beta \in \B\,, \forall \map A, \map B \in  \Act$, and {(ii)} $\map I \beta  =  \beta\, , \forall \beta \in \B$, where $\map I$ is the identity action. 


{\em Backward consistency.---}%
When actions are included in the picture,  a new class of \emph{sequential experiments} arise.  A sequential experiment consists of a sequence of actions interspersed by principal experiments. For example, $(\map A, \Part E, \map B, \Part F)$ 
represents a sequence consisting of an action $\map A$, followed by a principal 
experiment $\Part E$, by another action $\map B$, and by another principal experiment $\Part F$. Crucially, the belief $\beta$ supplies the agent with a joint probability distribution $
P_{  \beta ,\map A, {\mathbb E},  \map B, { \mathbb F} }  (E,F):= p( F|\map B E \map A \beta) \, p( E|\map A\beta)$ if $p( E|\map A\beta)\neq 0$, and zero otherwise.

When  the marginal probability  
${P }_{\beta, \map A, {\mathbb E},  \map B, { \mathbb F} }  (F) :  =    \sum_{E \in\mathbb E}  P_{\beta, \map A, {\mathbb E},  \map B, { \mathbb F} }(E,F)$
 is non-zero, one can define  the  conditional probability  
 $P_{\beta ,\map A, {\mathbb E},  \map B, { \mathbb F}} (E|F)  :=   P_{\beta, \map A, {\mathbb E},  \map B, { \mathbb F} }  (E,F)/P_{\beta, \map A, {\mathbb E},  \map B, { \mathbb F} } (F)$, which  quantifies the agent's confidence in \emph{retrodicting} that the earlier outcome must have been $E$, given   that the later outcome is $F$.     We say that the event $F$ is {\em unaffected by the experiment $\mathbb E$} if  the probability of $F$   in the  sequential experiment  $(\map A,  {\mathbb E},  \map B, {\mathbb F})$  coincides with the probability of $F$ in the sequential experiment  $ (\map A,   \map B, {\mathbb F})$.   In formula, $P_{\beta, \map A, {\mathbb E},  \map B, { \mathbb F}}   (F)=  P _{\beta, \map A,  \map B, { \mathbb F}}  (F)  :=  p(  F |\map B\map A \beta)$.

 If   $ P_{\beta, \map A, {\mathbb E},  \map B, { \mathbb F}} (E|F)=1$ and $F$ is unaffected by the experiment $\mathbb E$, we say that the event $F$ {\em implies} the event $E$.  In this case,  it is natural to require that  the update for event $E$  it is already included in  the update for event $F$.  
This idea motivates the following axiom:

\begin{ax}[Backward consistency]\label{ax:backward} 
If the event $F\in \mathbb F$ implies the event $E\in  \mathbb  E$  in a sequential experiment $(\map A, \Part E, \map B, \Part F)$, then the update for event $E$ can be  omitted in the final belief updated for event $F$. 
  Mathematically: if    $
P_{\beta,\map A, {\mathbb E},  \map B, { \mathbb F}} (E|F)=1$  and $P_{\beta,\map A, {\mathbb E},  \map B, { \mathbb F}}  ( F)   =  P_{\beta,\map A,\map B, { \mathbb F}}(  F)$, then $F \map B E \map A \beta = F \map B \map A \beta $.\end{ax}

Axioms \ref{ax:condprob}, \ref{ax:forward}, and \ref{ax:backward} define a set of theories, which we call {\em general Bayesian theories (GBTs)}.   In  Appendix \ref{app:GBTQ} we show that quantum theory is an example of GBT, with the principal experiments corresponding to projective measurements on a fixed standard basis
 and the updates following L\"uders' rule. 




{\em Ideal experiments.---}%
We now show that every GBT  contains a special class of ideal experiments that leave the 
agent with the option of gathering more refined  information in the 
future. 

We say that an experiment $(\map B,\Part F)$ with partition $\Part F= 
(F_{i,l})_{i,l}$ is a {\em refinement} of another experiment $(\map A, \Part 
E)$ with partition $\Part E= (E_i)_i$ if 
\begin{equation}
 \sum_l \, p( F_{i,l} | \map B \beta)= p( E_i |\map A \beta)\qquad \forall i \, , \forall \beta\in \B \, .
\end{equation}

The experiment $(\map A,\Part E)$ is {\em sequentially 
refinable} if there exists an action $\map A'$ such that, for every refinement 
$(\map B, \Part F)$ and for every initial belief $\beta$, the probability of 
the event $F_{i,l}$ in the experiment $(\map B, \Part F)$ is equal to the joint 
probability of the events $(E_i,F_{i,l})$ in the sequential experiment $ (\map 
A, \Part E,\map A', \map B, \Part F)$. In formula,
\begin{equation}\label{refinability}
 p(F_{i,l} |\map B \beta) = p( F_{i,l} | \map B \map A' E_i \map A \beta) \, 
 p( E_i | \map A\beta )
\end{equation}
for every refinement $(\map B, \Part F)$, for every event $F_{i,l} \in \Part 
F$, and for every belief $\beta \in \B$ with $p(E_i|\map A\beta)\ne 0$. In other words,   the coarse-grained experiment $(\map A, \Part E)$ does not alter the 
probability assignment for the fine-grained experiment $(\map B, \Part F)$, 
provided that the agent performs the  action $\map A'$ between them. Intuitively, if an experiment does not disturb any of its refinements then this 
experiment alters the belief as little as possible, this being  a 
central property of projective measurements  in quantum theory \cite{CY14, 
CY16,MK14}.

Consider now the  
family of experiments of the form $(\map A, \Part E)$ where the action $\map A$ 
is fixed and the partition $\Part E$ is variable. If there exists an action $\map A'$ such that condition \eqref{refinability} holds for 
all partitions $\Part E$, then we call each experiment $(\map A,\Part E)$ 
{\em ideal}. Such experiments exist in every theory: 
\begin{theo}
Ideal experiments exist in every GBT. In particular, every principal experiment $\Part E$ is ideal.
\end{theo}
The theorem is proved in Appendix  \ref{app:refinability}.  In Appendix \ref{app:projective}, we prove that the set of  ideal experiments in quantum theory coincides with the set of  projective measurements.

\begin{figure}
	\includegraphics[scale=0.2]{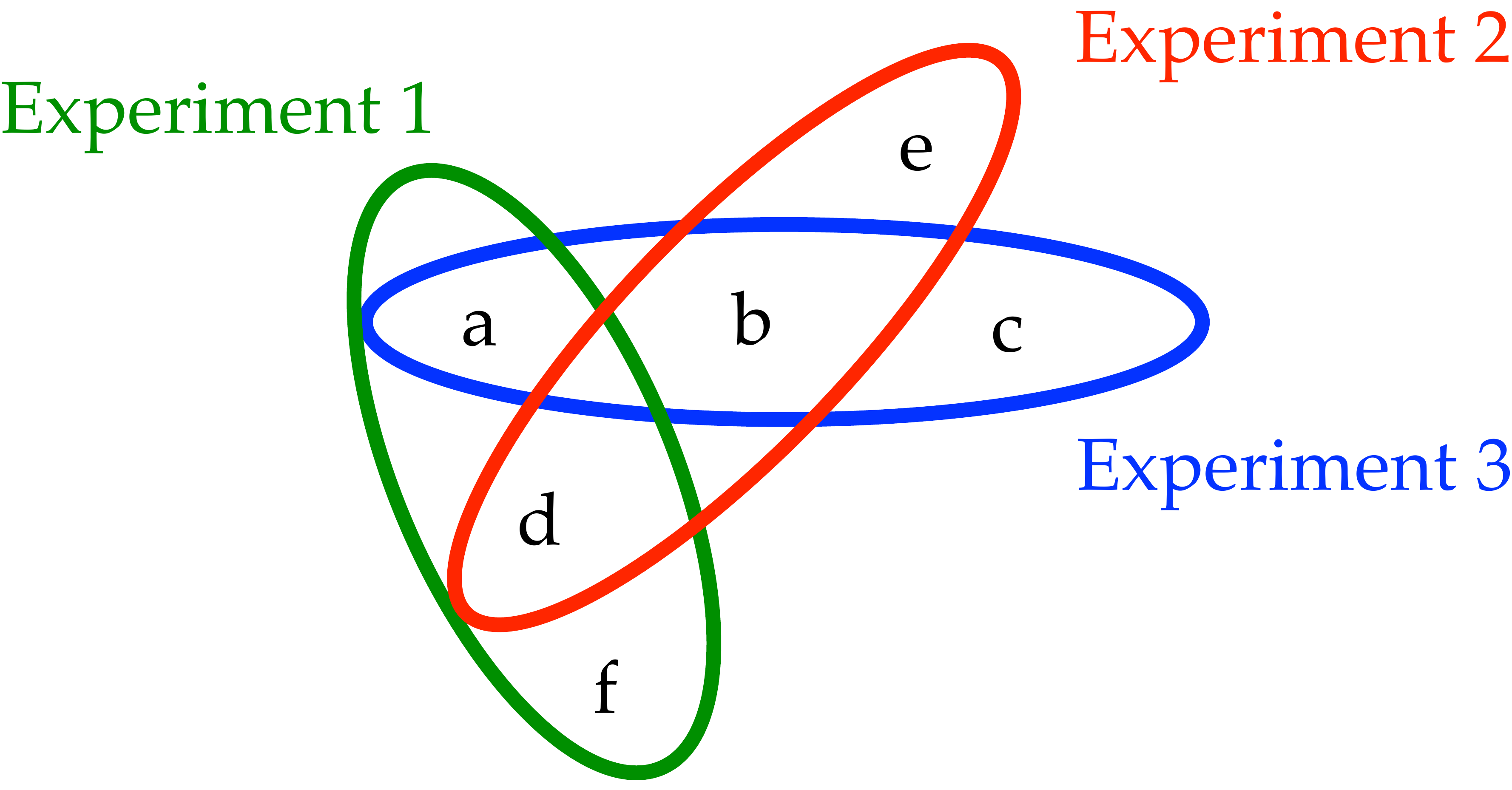}
	\caption{{Exclusivity principle in an example.}  Three experiments have possible outcomes ${\sf O}_1  = \{a,d,f\}$,  ${\sf O}_2 =\{ d,b,e\}$, and ${\sf O}_3  = \{a,b,c\}$, respectively. Each outcome $x$ has  a probability $p(x)$ assigned to it, and   the probabilities of the outcomes within each experiment sum up to  1, corresponding to the normalization condition $p(a)+ p(d) +  p(f) = p(d)+p(b) +p(e) =  p(a) +  p(b)+ p(c) =1$. 
	The outcomes $\{a,b,d\}$ satisfy the conditions  $\{a,b\}  \subset {\sf O}_3$,   $\{a,d\}  \subset {\sf O}_1$, and $\{b,d\}  \subset {\sf O}_2$, and therefore are called pairwise exclusive. The exclusivity principle demands $p(a)+  p(b) +  p(d) \le 1$. 
}
 \label{fig:EP} 
\end{figure}

{\em The emergence of the exclusivity principle.---}  The exclusivity principle, as originally introduced in the literature \cite{CSW10, Cabello13, Yan13,ATC14, 
Cabello15, Henson15, AFLS15},  refers to  scenarios where multiple alternative experiments  share some of their outcome, as in the example of 
 Figure \ref{fig:EP}.  Let $\cal S$ be the set of experiments under consideration, and, for each experiment $\cal E  \in \cal S$, let ${O}_{\cal E}$ be the set of its possible  outcomes. The outcomes of all experiments are assigned probabilities, with the constraint that the probabilities should sum up to 1 for every experiment, namely $\sum_{x  \in {O}_{\cal E} }  p(x)  =1, \, \forall \cal E  \in \cal S$.   A set of outcomes $O$, possibly belonging to different experiments, is called {\em pairwise exclusive} if, for every pair of outcomes $\{x_1,  x_2\}  \subseteq  O$ there exist an experiment ${\cal E} \in \cal S$ such that both $x_1$ and $x_2$ belong to ${O}_{\cal E}$.    A set of  experiments $\cal S$    satisfies the exclusivity principle if the condition  $\sum_{x  \in  O} p(x) \leq 1$ holds  for every set $O$ of pairwise exclusive outcomes. This is a non-trivial property, and  is not satisfied by all  probability assignments. 
In the example of Figure \ref{fig:EP},  the assignment $p(a)  = 1/2$,  $p(b)  =p(d)=   3/8$, $p(c) =p(f) = 1/8$, $p(e)=1/4$ satisfies the  condition  $p(a)+ p(d) +  p(f) = p(d)+p(b) +p(e) =  p(a) +  p(b)+ p(c) =1$ but violates the exclusivity principle.     

  We now show how to translate the exclusivity principle  in the language of GBTs.  The  principle applies to outcomes of ideal experiments.   By an outcome of an ideal experiment $\cal E    =(\map A,\mathbb{E})$ we mean a pair $x=  (\map A,  E)$ with $E  \in\mathbb E$.   The set of outcomes of the experiment $\cal E$ is then $O_{\cal E}    :=  \{    (\map A, E)~|~  E\in \mathbb E\}$.   
  
 To say that two experiments share an outcome, as in Figure  \ref{fig:EP},  we need a criterion to identify outcomes of different experiments. For two  experiments $ \cal E  = (\map A,  \mathbb E)$ and $ \cal E' = (\map A',  \mathbb E')$, we say that  two outcomes $ (\map A,  E)  \in  O_{\cal E}$ and $ (\map A',  E')  \in  O_{\cal E'}$   are  {\em equivalent}, denoted $ (\map A,  E) \simeq (\map A',  E')$,  if they have the same probability for every possible belief, namely $p(  E  |\map A  \beta)   =  p(E'  |\map A'  \beta)\, , \forall \beta\in\B$.   Diagrams like the one in   Figure \ref{fig:EP} arise when equivalent outcomes are identified.

Two outcomes $ (\map A,  E)  \in  O_{\cal E}$ and $ (\map A',  E')  \in  O_{\cal E'}$, possibly arising from different experiments $\map E \not  =\map E'$,  are mutually exclusive if  $(\map A,  E)$ and $(\map A',  E') $ are equivalent to two distinct outcomes of a single experiment $\cal F$, namely $(\map A,  E)\simeq x  $ and $(\map A',  E')\simeq x'$ for $\{x,  x' \}  \subseteq   O_{\cal F}$ and $x\not  =  x'$. 
A set of outcomes $O $ (of generally  different experiments) is {\em pairwise exclusive} if, for every pair of  outcomes $\{x,x'\} \subset   O$,   $x$ and $x'$ are  mutually exclusive.

We are now ready to translate the exclusivity principle in the framework of general Bayesian theories. The principle states that for every set  $O = \{  (\map A_n,  E_n)\}$ of pairwise exclusive outcomes of ideal experiments, the condition 
\begin{equation}\label{EP}
 \sum_{n} p(E_n|\map A_n  \beta) \le 1
\end{equation}
must be satisfied for every belief $\beta \in \B$. 

The central result of our paper is that the  bound (\ref{EP}) holds in every GBT: 
\begin{theo}
In every GBT, the outcomes of ideal experiments  satisfy the exclusivity principle.
\end{theo}
The proof is based on the refinability property of ideal experiments, which allows us to construct a sequence of binary experiments with the property that a subset of outcomes is equivalent to  the original set of pairwise exclusive outcomes.  Since all the outcomes in the pairwise exclusive set can be turned into outcomes of a single measurement, the bound (\ref{EP}) then follows from the normalization of the probability distribution for that measurement. The  details of the proof are provided in Appendix  \ref{app:specker}.   

Combining our derivation of the exclusivity principle with the result of Ref. \cite{Cabello18}, we obtain that the set of quantum correlations can be characterized completely in terms of Bayesian consistency conditions: 
\begin{cor}
In every GBT where every 
two statistically independent ideal experiments can be performed in parallel,  the largest set of correlations arising from  ideal experiments coincides with the quantum set for every Bell 
or Kochen--Specker contextuality scenario.
\end{cor}

The proof of Corollary~1  follows from a proof in \cite{Cabello18}, by replacing Lemma~1 in 
\cite{Cabello18} with our Theorem~2 and Assumption~2 in \cite{Cabello18} with the 
assumption that any two experiments can be performed in parallel with statistically independent joint distribution.


{\em Conclusion.---}%
No interpretation of quantum theory can be considered satisfactory unless one 
can derive the formalism of the theory from the key elements of the 
interpretation \cite{Fuchs}. This problem is especially pressing for QBism, which views quantum theory as a set of  rules that constrain how agents make bets about the outcomes of their 
experiments.  Can the quantum formalism be 
obtained from Bayesian consistency conditions alone? 

To address this question, we have introduced a  
framework for general Bayesian theories. 
The core of our framework  are three  consistency conditions on how the agent updates its beliefs:  the law of 
conditional probabilities, forward consistency, and 
backward consistency. These  conditions are sufficient to prove the existence of a set of  ideal 
experiments, which in quantum theory coincide with  the projective measurements.  We showed that 
the correlations arising from  ideal experiments must satisfy the exclusivity 
principle, which implies that the set of correlations  for every  Bell or Kochen-Specker 
contextuality scenario is equal to the corresponding set in quantum theory, provided that any two 
statistically independent ideal experiments admit a joint realization. 

Our result shows that the set of quantum correlations in Bell and Kochen-Specker contextuality scenarios can be derived from 
elementary consistency rules on how an agent should bet on the outcomes of 
future experiments. The problem of whether the whole quantum formalism can be 
recovered solely from Bayesian consistency conditions is still open. Unlike 
earlier reconstructions of quantum theory~\cite{Hardy01, CDP11, DB11, MM11, 
Hardy11, Wilce, HW} that adopted an interpretation agnostic approach, a fully Bayesian 
reconstruction  would have the 
potential to shed new light on the interpretation of quantum theory.


{\em Acknowledgments.---}
We thank Debbie Leung, Hans J. Briegel, \v{C}aslav Brukner, Christopher A. Fuchs, R\"udiger Schack, and Matthew Pusey for discussions during the development of this work.
This work is supported by
the Foundational Questions Institute through grant FQXi-RFP3-1325, the National Natural Science Foundation of China through grant 11675136, the Hong Kong Research Grant Council through grant 17300317, 
the Spanish Ministry of Science, Innovation and Universities (MICINN),
the Spanish Ministry of Economy, Industry and Competitiveness (MINECO),
the European Regional Development Fund FEDER through Grants No.\ FIS2017-89609-P and No.\ FIS2015-67161-P,
the ERC (Starting Grant No.\ 258647/GEDENTQOPT and Consolidator Grant 683107/TempoQ),
and the Basque Government (Grant No.\ IT986-16).
This research was supported in part by Perimeter Institute for Theoretical 
Physics. Research at Perimeter Institute is supported by the Government of 
Canada through the Department of Innovation, Science and Economic Development 
Canada and by the Province of Ontario through the Ministry of Research, 
Innovation and Science.


\newcounter{bayes}\setcounter{bayes}{1}
\newcounter{refinability}\setcounter{refinability}{3}
\newcounter{thm1}\setcounter{thm1}{1}
\newcounter{thm2}\setcounter{thm2}{2}
\newcounter{ax1}\setcounter{ax1}{1}

\appendix 

\section{Quantum theory as a general Bayesian theory}
\label{app:GBTQ}

Here we specify how quantum theory fits in  the framework of general Bayesian theories.

\subsection{Principal experiments, beliefs, and updates}  

For a quantum system with Hilbert space ${\cal H}  =  {\C}^d$, the sample space is the set $\Chi  = \{1,\dots, d\}$, and the possible events are subsets of $\Chi$.  

Any given partition of $\Chi$ into disjoint subsets, denoted by ${\mathbb E}  =(E_i)_i$,  is associated to a projective measurement with  orthogonal projectors $\{P_{E_i}\}$, defined as 
\begin{align}P_{E_i}  :=  \sum_{j \in  E_i}  \,  |j\>\<j| \, ,
\end{align} 
where $\{  |j\>\}$ is a fixed orthonormal basis for the Hilbert space $\cal H$.  In other words, the principal experiments are projective measurements that are diagonal in a fixed basis.   We call these measurements the {\em principal measurements}.   

 The set of beliefs is the set of density operators $\B :  =  \{  \rho  \in  L({\cal H}) ~|~   \rho \ge 0  \, , \Tr(\rho)=1\}$.       
For the principal experiment  ${\mathbb E}  =(E_i)_i$, the probability assigned to outcome $i$  is 
\begin{align}\label{pEi}
p(E_i  |  \rho)  =  \Tr (P_{E_i} \rho)\, ,
\end{align} where  $\rho  \in  \B$ is the initial belief.   If the outcome $i$ is guaranteed to be the case, and if $p( E_i |\rho) \not =  0$, then the belief $\rho$ is updated according to L\"uders' rule 
\begin{align}
\label{updatedEi} 
E_i \rho  &=    \frac{P_{E_i}\rho P_{E_i}}{\Tr(P_{E_i}\, \rho)} \,.
\end{align}
We call the non-linear map 
\begin{align}\label{updatemap}
\map M_{E_i} :\quad  \rho  \mapsto \map M_{E_i}  (\rho)  : = E_i  \rho
\end{align} the {\em update map}.  Note that the update map is only defined for $\Tr(  P_{E_i}  \, \rho)  \not  = 0$.

Note that Axiom 1 holds:  for every pair of events  $E \in  \Chi$ and $F\in \Chi$ such that $p(E|\rho ) \not  =0$, one has 
\begin{align}
\nonumber p(  F  | E  \rho)    & =    \Tr [P_{F}  \,    \map M_{E} (\rho)] \\
\nonumber & =    \frac{\Tr  (  P_F  \,   P_{E}\rho P_{E} )}{\Tr(P_{E}\, \rho)}  \\
\nonumber & =    \frac{\Tr  (   P_{E}   \, P_F  \,   P_{E}\rho  )}{\Tr(P_{E}\, \rho)}  \\
\nonumber &    = \frac{\Tr(P_{E \cap F}\rho )}{\Tr(P_{E} \rho)}\\
\nonumber & =  \frac{p(   E\cap F|\rho )}{  p(  E|\rho)} \, , 
\end{align}  
where the first equality follows from Eqs.~(\ref{pEi}) and~(\ref{updatemap}) with $E_i  = E$, the second equality follows from  Eq.  (\ref{updatedEi}) with $E_i  = E$, the third equality follows from  the ciclic property of  the trace,  the fourth equality follows from the relation $P_{E}  P_F  P_E  =  P_{E\cap F}$, and the last equality follows from Eq. (\ref{pEi}), applied  with $E_i  = E$ and with  $E_i  = E \cap F$. 

  \subsection{Actions and composite experiments} 
  
  The set of actions $\Act$ is the set of all possible quantum channels (completely positive trace-preserving maps)  from $ L  ({\cal H})$ to itself. The action of a channel $\map A $ on an input state $\rho$ can be written in the Kraus representation   $\map A (\rho)    =  \sum_i   A_i \rho  A_i$, where $(A_i)_i \subset  L(\cal H)$ is a set of Kraus operators, satisfying the normalization condition $\sum_i  A_i^\dag  A_i  = I$.     The quantum channels form a monoid, with the identity channel $\map I  :  \rho  \mapsto  \map I(\rho)  = \rho$ being the identity element.

   A composite experiment $(\map A,  \mathbb E)$ is described by a  set of completely positive trace non-increasing maps $(\map A_i)_i$, defined by  the relation $\map A_i (\rho)  :=   P_{E_i}  \map A (\rho)  P_{E_i}$, $\forall \rho$. The probability assigned to the outcome $i$ is  
  \begin{align}\label{borninstrument}
  p(E_i |\map A \rho)  =   \Tr[\map A_i (\rho)]   =:  p(\map A_i|\rho) \, ,
  \end{align} 
  and the updated state is  
  \begin{align}
\rho_i'   =    \frac{\map A_i (\rho)}{\Tr[\map A_i(\rho)]} \, .
  \end{align}
  
  Note that the sum $\sum_i \map A_i$ is a trace-preserving map, as it follows from Equation (\ref{borninstrument}).
  In general, a set of completely positive maps $(\map A_i)_i$ such that $\sum_i \map A_i$ is trace-preserving is called a {\em quantum instrument}.

 The set of actions includes all unitary channels $\map U$ of the form $\map U(\rho)  =  U\rho U^\dag$, where $U$ is a generic unitary operator.   Unitary channels represent the reversible actions in quantum theory.  
     Combined with the principal measurements, the unitary channels generate all projective measurements on the given quantum system.     
      
   

\subsection{Forward consistency}   Let $\rho$ be a density matrix, $E  \subseteq \{  1,\dots, d\}$ be an event, $P_E  =  \sum_{j  \in  E}    |j\>\<j|$ be the corresponding projector, and $\overline P_E:= I-P_E$ be its orthogonal complement.  The condition that 
the event  $E$ happens with certainty on $\rho$ is $\Tr(P_E \rho) =1$. 
Forward consistency is the statement that, under this condition, the updated 
state $\map M_E (\rho)  = P_E \rho P_E / \Tr(P_E \rho) $ is equal to the original state $\rho$. 

The proof 
is simple. The certainty condition $\Tr(P_E\rho) =1$ is equivalent to $\Tr 
(\overline P_E\rho \overline P_E )=0$. Since this equation is of the form $\Tr(AA^\dag) = 0$ with 
$A=\overline P_E\sqrt \rho$, we have $A=0$ and thus $\overline P_E\sqrt\rho = 0$ and $\sqrt \rho \, \overline P_E = 0$. This leads to
\begin{align}
\nonumber  \rho  & = ( P_E+ \overline P_E) \rho ( P_E + \overline P_E ) \\
\nonumber &= P_E\rho P_E\\
&  =P_E\rho P_E/\Tr(P_E\rho)  ,
\end{align}
the last equality following from the certainty condition $\Tr(P_E \rho)=1$. Since   $P_E\rho P_E/\Tr(P_E\rho)$  is the updated state through event $E$  [cf. Equation (\ref{updatedEi})], the above relation can be rewritten as $\rho  =  E \rho$. Hence,  quantum theory satisfies forward consistency.

\medskip 

\subsection{Backward consistency}
Backward consistency refers to sequential experiments of the form $(\map A, \Part E, \map B, \Part F)$, where $\map A$ and $\map B$ are two actions, and $\Part E$ and  $\Part F$ are two partitions of the sample space $\Chi$.  In the quantum case, the two actions $\map A$ and $\map B$ are quantum channels, and $\Chi  = \{1,\dots, d\}$.

 For a given density matrix $\rho$,    the  joint probability distribution of the events $E_i  \in  \Part E$ and $F_j  \in \Part F$ is  
 \begin{align}
  P_{\rho, \map A,{\mathbb E},  \map B,  {\mathbb F}}  (E_i,F_j)     :=  p\Big(  F_j  \Big|    (\map B   \circ \map M_{E_i} \circ\map A) (  \rho)  \,  \Big)\,  p\Big( E_i \Big|\map A    (\rho )\, \Big)
  \end{align}
for $p\big( E_i  |\map A  (\rho)  \, \big)  \not=  0$,  and ${\sf Prob} (E_i,F_j)   = 0$ otherwise. 

Using Eqs.~(\ref{pEi}) and (\ref{updatedEi}),  the joint probability distribution can be expressed as  
\begin{align}\label{easyjoint}
    P_{\rho, \map A,{\mathbb E},  \map B,  {\mathbb F}}  (E_i,F_j)  =  \Tr\left[ P_{F_j}  \map B  \Big(  P_{E_i}  \,\map A(\rho) \,  P_{E_i}  \big) \right]  \, ,
\end{align}
both for  $p\big( E_i  |\map A  (\rho)  \, \big)  \not=  0$ and for  $p\big( E_i  |\map A  (\rho)  \, \big)  =  0$.

Let $E  =  E_{i_0}$ and $F=  F_{j_0}$ be two given events in $\mathbb E$ and $\mathbb F$, respectively.  
In the following, we will  show that $  P_{\rho, \map A,{\mathbb E},  \map B,  {\mathbb F}}  (E_i,F_j) ( E|F) =1$  implies the condition 
\begin{align}\label{desideratum}
 (  \map M_F \circ\map B \circ  \map M_E\circ  \map A ) \, (\rho)    =  ( \map M_F \circ \map B \circ \map A  ) \, (\rho)  \, .  
\end{align}  
By proving this condition, we will prove that quantum theory satisfies  backward consistency (in fact, we will prove a slightly stronger result, since the condition (\ref{desideratum}) holds without the need of imposing that the event $F$ is unaffected by the experiment $\mathbb E$).

Let us prove condition (\ref{desideratum}). For brevity, we define $\rho' :  =  \map A  (\rho)$, so that Eq.~(\ref{easyjoint}) becomes
\begin{align}
P_{\rho, \map A,{\mathbb E},  \map B,  {\mathbb F}}  (E_i,F_j)      =  \Tr[ P_{F_j}  \map B  (  P_{E_i}  \,\rho' \,  P_{E_i}  ) ]  \, , \label{easyjoint1}
\end{align}
  
 The condition $P_{\rho, \map A,{\mathbb E},  \map B,  {\mathbb F}} ( E|F) =1$ is equivalent to the condition $P_{\rho, \map A,{\mathbb E},  \map B,  {\mathbb F}} (E_i  | F) =  0$ for every $i\not = i_0$.  Hence, we also have  $P_{\rho, \map A,{\mathbb E},  \map B,  {\mathbb F}} (E_i  ,  F)  =  0$ for every  $i\not = i_0$, namely   
\begin{equation}\label{B3}
 \Tr[ P_F \map B ( P_{E_i} \rho'   P_{E_i}) ] =0 \qquad \forall i\not = i_0 \, .
\end{equation}
Using a Kraus representation  $\map B  (\cdot)  =  \sum_j   B_j  \cdot  B_j^\dag$ and the relation $P_E  =  P_E^2$, we obtain  
\begin{align}
\sum_j   \Tr  \left[  (P_F   B_j P_{E_i} \sqrt{\rho'})~( \sqrt{ \rho'} P_{E_i} B_j^\dag P_F)\right]= 0 \qquad\forall i\not = i_0  \, ,
\end{align}
which implies  
\begin{align}
P_F B_j P_{E_i}\sqrt\rho'= 0  \qquad \forall   j , \,  \forall  i\not = i_0 
\end{align}
Using this relation, we obtain 
\begin{align}
 \nonumber P_F\map B(\rho') P_F
 \nonumber &= \sum_{j,i,i'}   (P_FB_j P_{E_i} \sqrt{\rho'} ) \,  (  \sqrt{ \rho'}  P_{E_{i'}} B_j^\dag P_F) \\
\nonumber  &= \sum_j P_FB_j P_{E_{i_0}}\rho' P_{E_{i_0}} B_j^\dag P_F \\
 &= P_F\map B(P_E\rho' P_E)  P_F,
\end{align}
or equivalently,  
\begin{align}\label{mnb}
(\map M_F  \circ \map B ) (\rho')    =     ( \map M_F   \circ \map B \circ \map M_E )  (\rho') \, .    
\end{align}
Recalling the definition $\rho': =  \map A (\rho)$, we then obtain the desired Eq.~(\ref{desideratum}).


\section{Proof of Theorem~\arabic{thm1}}
\label{app:refinability}


In this section, we adopt the following definition: 
\begin{defi} 
An action $\map A$ is \emph{ideal} if the experiment $(\map A,\mathbb{E})$ is ideal for every partition $\mathbb{E}$.
\end{defi}

Using this definition, Theorem~\arabic{thm1} of the main text can be reformulated as 
\begin{theobis}\label{theobis}
 The identity action is ideal. 
\end{theobis}

To prove this statement, we will  use the following lemma:

\begin{lem}
Let $(\map C, \Part F)$ be a refinement of $(\map I, \Part E)$, with $\Part F= 
(F_{i,l})_{i,l}$ and $\Part E= (E_i)_i$, and let $F_i$ be the event defined by 
$ F_i: = \bigcup_l F_{i,l}$. Then,
\begin{equation}\label{refinement1}
 p(F_i|\map C\beta)=p(E_i|\beta) \qquad \forall i\,,\forall \beta\in \B
\end{equation}
and the equality
\begin{equation}\label{commute}
F_i \map C \beta = \map C E_i \beta
\end{equation}
holds  for every belief $\beta \in B$ and for every for every outcome $i$ such that $p(E_i|\beta) 
\ne 0$. \end{lem} 

{\bf Proof.}
Since $(\map C, \Part F)$ is a refinement of $( \map I, \Part E)$, 
one has
\begin{equation}\label{refine}
 p(F_{i} | \map C \beta) = \sum_l p(F_{i,l} | \map C \beta) = p( E_i |\beta)\,,
\end{equation}
where the first equality follows from the additivity of the probabilities of disjoint 
events.  Hence, Eq.~\eqref{refinement1} holds. 

Now, suppose that one has $p(E_i|\beta) 
\ne 0$.  Applying Eq.~\eqref{refinement1} to the initial belief $E_i\beta$, we obtain 
the condition
\begin{equation}\label{equalto1}
 p( F_i |\map C E_i \beta) = p( E_i |E_i \beta) = \frac{p(E_i\cap 
 E_i|\beta)}{p(E_i|\beta)}= 1,
\end{equation}
the second equality following from Eq.~(\arabic{bayes}) in the main text. Since 
the agent is certain that the event $F_i$ will occur, forward consistency implies 
that no update should take place on the belief $\map C E_i \beta$. Hence, we 
obtain the relation
\begin{equation}\label{uno}
 F_i \map C E_i \beta = \map C E_i \beta\,.
\end{equation}

The next step is to apply backward consistency. Consider the joint probability distribution   $P_{\beta ,  {\mathbb E},  \map C  ,{\mathbb F}}  (E_j,  F_k)  :=p( F_{k} | \map C E_j \beta ) 
\, p(E_j|\beta)$.  For every pair $(E_j, F_k)$ such that  $ p (E_j|\beta )  \not = 0$, one has 
\begin{align}
\nonumber P_{\beta ,  {\mathbb E},  \map C , {\mathbb F}} (E_j, F_k)    &   =  p(  F_k |  \map C E_j  \beta)  \,   p(E_j|\beta)  \\
   \nonumber & = p(  E_k |   E_j  \beta)  \,   p(E_j|\beta)   \\
\nonumber   &=     p(  E_k \cap   E_j  |\beta)\\  &= \delta_{j,k}  ~  p(  E_k | \beta) \, ,  \label{correlated}
\end{align}
where the second equality follows from Eq.~(\ref{refine}),  the third equality follows from the rule of conditional probabilities (Axiom \arabic{ax1} in the main text), the fourth equality follows from the fact that the events $(E_j)_j$ are mutually disjoint. 
 
 Eq.~(\ref{correlated}) implies that the experiment $\mathbb E$ does not affect the probability of the event $F_i$: 
 \begin{align}
\nonumber  P_{\beta ,  {\mathbb E},  \map C , {\mathbb F}}  (F_i)    & :  =   \sum_j P_{\beta ,  {\mathbb E},  \map C,  {\mathbb F}} (E_j, F_i)   \\
\nonumber &  =    \sum_{j:  P_{\beta ,   {\mathbb E},  \map C  {\mathbb F}} (E_j, F_i)
 \not = 0  }  P_{\beta ,  {\mathbb E},  \map C , {\mathbb F}} (E_j, F_i)\\
\nonumber &  =    \sum_{j:  p(E_j  |\beta) \not = 0  }  P_{\beta ,  {\mathbb E},  \map C,  {\mathbb F}} (E_j, F_i)\\ 
\nonumber &=  p(E_i| \beta)  \\
\nonumber & =  p(F_i|\map C \beta)\\
&  =   P_{\beta ,   \map C , {\mathbb F}}    \, ,\label{unaffected}
 \end{align}      
where the fourth equality follows from Eq.~(\ref{correlated}), and the fifth equality follows from Eq.~(\ref{refine}).  

Moreover, Eq.~(\ref{correlated}) implies   the condition $ P_{\beta ,  {\mathbb E},  \map C,  {\mathbb F}} (E_i|F_i)  = 1$. Note that the conditional probability is well defined because the marginal $P_{\beta ,  {\mathbb E},  \map C , {\mathbb F}}  (F_i)$ is non-zero.  Indeed, Eq.~(\ref{unaffected}) implies  the condition $P_{\beta ,  {\mathbb E},  \map C , {\mathbb F}}  (F_i)   =  p( F_i|\map C  \beta)  =  p(  E_i|\beta)  \not  =  0$.    

Hence, the hypotheses for the applications of backward consistency are satisfied.   By applying  backward consistency, we obtain the relation 
\begin{equation}\label{due}
 F_i \map C E_i \beta = F_i \map C \beta \,.
\end{equation}
Combining Eqs.~\eqref{uno} and~\eqref{due}, we obtain Eq.~\eqref{commute}. \qed 

\medskip 

We are now ready to provide the proof of Theorem \ref{theobis}.

{\bf Proof of Theorem \ref{theobis}.}
For an arbitrary partition $\Part E = (E_i)_i$ we consider an arbitrary 
refinement $(\map C,\Part F)$ of $(\map I, \Part E)$ with $\Part 
F=(F_{i,l})_{i,l}$. Let $F_i$ be the event defined by $ F_i: = \bigcup_l 
F_{i,l}$. Using the rule of conditional probabilities, we obtain for 
$p(F_i|\map C\beta)\ne 0$,
\begin{equation}
\label{provedseqref}
\begin{split}
 p( F_{i,l} | \map C \beta) & = p(F_{i,l} \cap F_i |\map C\beta) \\
 &= p ( F_{i,l} | F_i \map C \beta) \, p( F_i |\map C \beta) \\
 &= p ( F_{i,l} | \map C E_i \beta) \, p( E_i | \beta)\,,
\end{split}\end{equation}
the last equality following from Eqs.~\eqref{commute} and \eqref{refinement1}. 
If $p(F_i|\map C\beta)=0$, then Eq.~\eqref{refinement1} implies $p(E_i|\beta)= 
0$. Eq.~\eqref{provedseqref} is exactly the condition for sequential 
refinability, Eq.~(\arabic{refinability}) from the main text with $\map A=\map 
A'=\map I$. Since these considerations hold for any partition $\Part E$ and any 
refinement of $(\map I,\Part E)$, the action $\map I$ is ideal. \qed
\medskip 

In addition, we now prove the a strengthening of Theorem~\ref{theobis}, extending its statement to arbitrary   
reversible actions.   Here, we call an  action $\map A$  {\em reversible} if there exists another action $\map A^{-1}$ that acts as its inverse, namely $\map A^{-1} \map A\beta = \map A \map A^{-1} \beta= \beta, \, \forall \beta\in \B$.

\begin{cor}
If an action $\map A$ is reversible, then $\map A$ 
is ideal. 
\end{cor} 

{\bf Proof.} We need to show that for every partition $\Part E = (E_i)_i$, for 
every refinement $(\map B,\Part F)$ of $(\map A, \Part E)$, and for every 
belief $\beta$, the condition of sequential refinability, 
Eq.~(\arabic{refinability}) from the main text, is satisfied.

The  condition that $(\map B, \Part F)$ is a refinement of 
$(\map A, \Part E)$ reads
\begin{equation}
 \sum_l \, p( F_{i,l} | \map B \beta)= p( E_i |\map A \beta)\qquad \forall i \, , \forall \beta\in \B \, .
\end{equation}

The condition of sequential refinability is that  there exists some suitable action $\map A'$ such that  
\begin{equation}\label{refinability}
 p(F_{i,l} |\map B \beta) = p( F_{i,l} | \map B \map A' E_i \map A \beta) \, 
 p( E_i | \map A\beta ) \, ,
\end{equation}
for every refinement $(\map B, \Part F)$  of 
$(\map A, \Part E)$, for every event $F_{i,l} \in \Part 
F$, and for every belief $\beta \in \B$ with $p(E_i|\map A\beta)\ne 0$.

Since $\map A$ is reversible, all elements of $\B$ can be obtained as $ \map 
A\beta$. Hence,  $\beta':= \map A \beta$ is a generic element of $\B$.   
Setting  $\map C := \map B \map A^{-1}$, the condition  that $(\map B, \Part F)$ is a refinement of 
$(\map A, \Part E)$ becomes 
\begin{equation}
\sum_l\, p(F_{i,l} | \map C \beta') = p( E_i |\beta')
\qquad \forall \beta' \in \B \,. 
\end{equation}
This condition is equivalent to the fact that $(\map C, \Part F)$ is a 
refinement of $(\map I, \Part E)$.  Since $\map I$ is an ideal action (Theorem~\ref{theobis}), the 
 experiment  $(\map I, \Part E)$ is ideal, and satisfies the sequential refinability condition  
 \begin{equation}\label{refinable2}
p( F_{i,l} | \map C \beta') = p(F_{i,l}| \map C E_i \beta') \,
p(E_i |\beta' )
\end{equation}
for every refinement $(\map C, \Part F)$  of 
$(\map I, \Part E)$, for every event $F_{i,l} \in \Part 
F$, and for every $\beta'\in \B$ with $p(E_i|\beta')\ne 0$.

Recalling the definitions $\beta'  :=  \map A \beta$ and $\map C: = \map B  \map A^{-1}$, we can express the above condition as 
\begin{equation}\label{refinable2}
p( F_{i,l} | \map B \map A^{-1}  \map A   \beta) = p(F_{i,l}| \map B \map A^{-1} E_i \map A\beta) \,
p(E_i |  \map A\beta )
\end{equation}
for every refinement $(\map C, \Part F)$  of 
$(\map I, \Part E)$, for every event $F_{i,l} \in \Part 
F$, and for every $\beta\in \B$ with $p(E_i|\map A\beta')\ne 0$.

Comparing Eq. (\ref{refinable2}) with Eq. (\ref{refinability}), we can see that Eq. (\ref{refinability}) holds with $\map A'  =  \map A^{-1}$. 

Hence the condition for sequential refinability is satisfied for $(\map A,\Part 
E)$ applied to the experiment $(\map B, \Part F)$. Since the above argument 
holds for every partition $\Part E$ and the reversing action is always $\map 
A'$, the action $\map A$ is ideal.\qed

\section{Ideal experiments in quantum theory are projective measurements}\label{app:projective} 

Here we show that our notion of  an ideal experiment coincides with the notion of a projective  measurement  in quantum theory. 

We consider general quantum  measurements, described by quantum instruments, that is, collections $(\map A_i)_i$ of completely positive trace non-increasing maps, with the property that the sum $  \sum_{i}  \map A_i$ is trace-preserving.  For a quantum system in the state $\rho$, the probability of the outcome $i$ is  
\begin{align}
\nonumber p(  \map A_i |  \rho)    &:=\Tr [\map A_i (\rho)] \\
  & = \Tr[ A_i  \rho]  \qquad A_i  :  =   \map A_i^\dag (I) \, ,
  \end{align}
where   $\map A_i^\dag$ is the adjoint of the map $\map A_i$, uniquely defined by the condition $\Tr[ O  \map A_i(\rho) ]  =   \Tr[  \map A_i^\dag  (O)  \,   \rho]$, for all operators $\rho$ and $O$. Note that the operators $( A_i)_i$  form a positive operator-valued measure (POVM), namely $A_i\ge 0$  and $ \sum_i A_i  =  I$, the second condition following from the fact that the map $\sum_i  \map A_i$ is trace-preserving.

A quantum instrument   $  (\map B_{i,j})_{i,j}$ is a refinement of the  quantum instrument $(\map A_i)_i$ if one has  
\begin{align}
p(  \map A_{i}  | \rho)      = \sum_j  \, p(  \map B_{i,j} |\rho) \qquad \forall i \, , \forall \rho \,,  
\end{align}
or equivalently, 
\begin{align}
\Tr (A_{i}   \rho)      = \sum_j  \, \Tr(  B_{i,j}   \rho) \qquad \forall i \, , \forall \rho  \,,  
\end{align}
with $ B_{i,j} :  =  \map B_{i,j}^\dag (I)$.   Since this equation must hold for every state $\rho$, it can be simplified as  
\begin{align}\label{refineAi}
A_i    =  \sum_j \,  B_{i,j}  \qquad \forall i,j \, .
\end{align} 

A quantum instrument is sequentially refinable if there exists a quantum channel $\map A'$ such that, for every refinement $  (\map B_{i,j})_{i,j}$, one has 
\begin{align}\label{seqrefAB}
p(  \map B_{ij} | \rho)   =    p( \map B_{ij}  |  \map A'   \rho_i') \,  p(\map A_i  |  \rho) \qquad \forall i, j  \, , \forall \rho \, ,
\end{align}
where  $\rho_i'$ is the conditional state $\rho_i'  :=  \map A_i (\rho)/\Tr[\map A  (\rho)]$.

With these definitions, we have the following lemma: 
\begin{lem}\label{lem:refinableprojective}
If a quantum instrument $(\map A_i)_i$ is sequentially refinable, then all POVM operators $A_i  :=  \map A_{i}^\dag (I)$ are projectors. 
\end{lem}

{\bf Proof.}  Equation (\ref{seqrefAB}) can be rewritten  as 
\begin{align}
\Tr (B_{i,j} \rho)    =     \Tr[B_{i,j}   \map A_i'  (\rho)  ] \qquad \forall i,j \, , \forall \rho ,
\end{align}
with $\map A_i'    :=    \map A'\map A_i$.  
Equivalently,  we have
\begin{align}\label{povmcondition}
B_{i,j}    =    \map A_i^{\prime \dag}    (  B_{i,j})  \qquad \forall i,j\, . 
\end{align}
 
 Now,  the above equation holds  for every POVM $(B_{i,j})_{i,j}$ that refines the POVM $(A_i)_i$, meaning  that   Eq. (\ref{refineAi}) holds.  Note that  every non-negative operator $B_i$ with support contained in the support of $A_i$ can be rescaled in such a way that $  B_i\le A_i$.  Hence, one can define a refinement $(B_{i,j})_{i,j}$  with $B_{i,0}  = B_i$ and $B_{i,1}  =  A_i- B_i$ and apply condition (\ref{povmcondition}) to it, thus obtaining 
\begin{align}\label{povmcondition1}
B_i    =  \map A_i^{\prime \dag}  (B_i) \qquad  \forall i, \forall B_i :   \Supp (B_i) \subseteq \Supp (A_i)\, ,
\end{align}
where $\Supp (A_i)$ and $\Supp (B_i)$ denote the supports of $A_i$ and $B_i$, respectively. 

On the other hand, the map $\map A_i^{\prime \dag}$ annihilates every operator $C$ with support in the orthogonal complement of $\Supp (A_i)$, namely
\begin{align}\label{annichilate}
\map A_i^{\prime \dag} (  C)   =  0  \qquad \forall i, \forall C :  \Supp (C)  \subseteq \big(\Supp (A_i)\big)^\perp \, .
\end{align} 
The proof of Equation (\ref{annichilate}) is as follows.  By definition, one has 
\begin{align}\label{lkjh}
\nonumber \sum_i  A_i   & =   I   \\
\nonumber     & =  \sum_i   \map A_i^{\prime \dag}   (I) \\
\nonumber  &  \ge  \sum_i   \map A_i^{\prime \dag} (A_i)  \\
&  =  \sum_i   A_i  \, ,   
\end{align}
the second equality following from the fact that the map $\sum_i  \map A_i^{\prime}$ is trace-preserving, and the last equality following from  Equation (\ref{povmcondition1}) in the special case $B_i=  A_i$.  In order for Equation (\ref{lkjh})  to hold,  we must have 
\begin{align}\label{ortho} 
\map A_i^{\prime \dag}  (I)   =  \map A_i^{\prime \dag}  (A_i)   \qquad \forall i \, .
\end{align}
 Now, every operator $C$ with support in the orthogonal complement  $\big(  \Supp (A_i)\big)^\perp $  is proportional to an operator $C'$ such that $C' \le  I- A_i$. Writing  $C$ as $C=  \lambda \, C'$ for some constant $\lambda >  0$, we have 
 \begin{align}
\nonumber \map A_i^{\prime \dag}  (C)  &   = \lambda^{-1}  \,    \map A_i^{\prime \dag}  (C')   \\
 \nonumber  &  \le  \lambda^{-1} \,   \map A_i^{\prime \dag} (I-A_i)  \\
 \label{tozero}   &  = 0 \, ,  \qquad \forall i
\end{align}  
which proves Equation~(\ref{annichilate}).

To conclude the proof, let us define  $\Pi_i$ to be the projector on the support of $A_i$.  Combining Equations~(\ref{annichilate}) and  (\ref{povmcondition1}), we obtain 
\begin{align}
\nonumber \map A_i^{\prime \dag}  (I)   &  =   \map A_i^{\prime \dag}   (  \Pi_i )  +   \map A_i^{\prime \dag}   (  I-  \Pi_i ) \\
& =    \Pi_i   \qquad \forall i \, .
\end{align}
Note that, by definition, one has $\map A_i^{\prime \dag}  (I)   =  \map A_i^\dag \map A^\dag (I)  =    \map A_i^\dag  (I)$,  because $\map A$ is trace-preserving and therefore $\map A^\dag (I)  = I$.   

Summarizing, we have shown that every operator $A_i  :  =  \map A_i^\dag  (I)$ is a projector.   
Moreover, the projectors $(A_i)_i$ must be orthogonal. Indeed, for every $i\not =  j$,  one has 
\begin{align}
 \nonumber \Tr(A_i) &  =      \sum_k  \Tr(A_i  A_k)\\
\nonumber  &\ge   \Tr (A_i^2)  +   \Tr(A_i A_j)  \\
\nonumber  &  \ge   \Tr(A_i) \, , \end{align} 
which implies  $\Tr(A_i  A_j)=  0$, or equivalently $\Tr[(   A_i A_j )  (A_j  A_i)]=0$.  Hence, one must have 
$A_i A_j  = 0$. 

In conclusion, the POVM operators $(A_i)_i$ are orthogonal projectors.   \qed 

\medskip

\begin{lem}\label{lem:refinablelueders}
If a quantum instrument $(\map A_i)_i$ is sequentially refinable and $\map A'$ is the quantum channel in  the refinability condition (\ref{seqrefAB}), then the quantum instrument defined by $\map A_i'  : = \map A' \map A_i$ has the L\"uders form $\map A_i' (\cdot)  =  A_i  \cdot A_i$ for every outcome $i$. 
\end{lem}

 {\bf Proof.} Using Equation~(\ref{annichilate}) with $C  = A_i^\perp  : =  I-A_i$, we obtain the relation  
 \begin{align}\label{rfv}
 \map A_i^{\prime  \dag}  (A_i^\perp)  =  0   \qquad \forall i  \, .
 \end{align}     
  Let $ \map A_i'  (\rho)  =  \sum_j  A_{i,j}' \rho  A_{i,j}^{\prime \dag}$ be a Kraus representation for $\map A_i'$. 
  Inserting it in Equation (\ref{rfv}) we obtain 
  \begin{align}
\sum_j  \Tr[  A_{i,j}^{\prime \dag}   A_i^\perp  A_{i,j}' ]   =  0    \qquad \forall i \, ,
  \end{align}
  and therefore 
   \begin{align}
 \Tr[  A_{i,j}^{\prime \dag}  A_i^\perp A_{i,j}' ]   =  0  \qquad \forall i,j \, .
  \end{align}
  Since $A_i^\perp$ is a projector, the above equation can be rewritten as 
  \begin{align}
 \Tr[    (A_{i,j}^{\prime \dag}  A_i^\perp   )(  A_i^\perp A_{i,j}' )]   =  0  \qquad \forall i,j \, ,
  \end{align}
 which implies $A_{i,j}^{\prime \dag}  A_i^\perp   =   A_i^\perp  A_{i,j}'  = 0$ for every $i$ and $j$.
 
For a generic operator $O$, we then have 
\begin{align}
\nonumber  \map A_i^{\prime \dag}  (O)    &   =  \sum_j    A_{i,j}^{\prime \dag}  \,  O   A_{i,j}'  \\
\nonumber &  =   \sum_j    A_{i,j}^{\prime \dag}  (A_i  +  A_i^\perp)  \,  O      (A_i  +  A_i^\perp)  A_{i,j}'  \\
\nonumber &  =   \sum_j    A_{i,j}^{\prime \dag}  A_i    O      A_i    A_{i,j}'  \\
 \nonumber &  =  \map A_i^{\prime \dag}   (  A_i  O A_i)  \\
 &  =  A_i  O  A_i  \qquad \forall i \, , 
\end{align}
   having used Equation (\ref{povmcondition1}) with $B_i   =  A_i  O A_i$.  
   \qed 
   
   \medskip 
   
In summary, ideal experiments in quantum mechanics are  projective measurements, with state update given by L\"uders rule. The converse is also true: 
\begin{lem}
Every projective measurement, with state update given by L\"uders' rule, is sequentially refinable.
\end{lem}

{\bf Proof.}  Let $(A_i)_i$ be a projective POVM ({\em i.e.} a set of projectors such that $\sum_i  A_i =  I$),  $(\map A_i)_i$ be the quantum instrument defined by $\map A_i(\rho)  :=  A_i \rho  A_i$, $\forall \rho$, and $( \map B_{ij})_{i,j}$ be a quantum instrument that refines $(A_i)_i$, meaning that the corresponding   POVM  $(B_{i,j})_{i,j}$, with $B_{i,j}:  =  \map B^\dag_{i,j}  (I)$, satisfies  Equation (\ref{refineAi}).    
 Then, one has 
 \begin{align}
 \nonumber p( \map B_{i,j}  |   \rho )  &  =  \Tr[  B_{i,j}  \rho ]\\
  \nonumber & =\Tr[  (A_i  B_{i,j}  A_i) \rho]\\
\nonumber  & =\Tr[B_{i,j}   (A_i  \rho  A_i)]\\
\nonumber  & =\Tr[B_{i,j}  \map A_i  (\rho)]  \\
 &  =  p(  \map B_{i,j}   |  \rho_i')  \,  p(  \map A_i  |\rho)  \qquad \forall \rho \, , 
 \end{align}
 with $\rho_i'   =  \map A_i(\rho)/\Tr[\map A_i(\rho)]$.   Hence, the sequential refinability condition (Equation 4 of the main text) holds with $\map A'  =  \map I$.  \qed 
 
 \medskip 
 
 In summary, we have proven the following theorem: 
 \begin{theo}
 The set of ideal experiments in quantum theory coincides with the set of  projective measurements, with state update given by L\"uders rule. 
 \end{theo}

\section{Proof of Theorem~\arabic{thm2}}
\label{app:specker}


The proof of Theorem~\arabic{thm2} follows the line of argument in 
Ref.~\cite{CY14}.

\begin{lem}\label{lem:sameeffect}
Let $(\map A_1,E_1)$ and $(\map A_2,E_2)$ be two mutually exclusive ideal 
outcomes and let $\map A_1'$ be the reversing action in condition (4) of the main text. Then, one has the following implications: 
\begin{enumerate}
\item  If $p(\overline{E}_1|\map A_1\beta)\ne 0$,  then
\begin{equation}\label{lem5}
 p(E_2| \map A_2 \map A'_1 \overline{E}_1\map A_1 \beta)\,
 p( \overline{E}_1|\map A_1\beta)= p(E_2| \map A_2\beta) \, .
\end{equation}
\item  If  $p(\overline{E}_1|\map A_1\beta)= 0$, then  $p(E_2|\map A_2\beta)   =0$.
\end{enumerate}
\end{lem}

{\bf Proof.} Since $(\map A_1,E_1)$ and $(\map A_2,E_2)$ are mutually 
exclusive, there exists an experiment $(\map B, \Part F)$ and two  events $F_1 ,  F_2\in \mathbb F$ satisfying $p(F_1|\map B\beta)= p(E_1|\map A_1\beta)$ and 
$p(F_2|\map B\beta)= p(E_2|\map A_2\beta)$ for all beliefs $\beta$.  The 
experiment $(\map B,\Part F)$ is a refinement of the experiment $(\map 
A_1,\Part E_1)$ with $\Part E_1=(E_1,\overline{E}_1)$. This holds since for any 
belief $\beta$ we have
\begin{equation}
p( F_1|\map B \beta) = p( E_1 |\map A_1 \beta)
\end{equation}
and
\begin{equation}\label{emab2}\begin{split}
\sum_{i\ne 1}p( F_i |\map B \beta ) &= 1- p(F_1|\map B\beta) \\
 & = 1- p(E_1|\map A_1\beta) = p( \overline{E}_1 |\map A_1 \beta) \,.
\end{split}\end{equation}
Now, there are two possibilities: either  $p(\overline{E}_1|\map A_1\beta)\ne 0$  or $p(\overline{E}_1|\map A_1\beta)= 0$. For  $p(\overline{E}_1|\map A_1\beta)\ne 0$,  the sequential refinability of the experiment $(\map A_1,\Part E_1)$ 
implies the equality
\begin{equation}\begin{split}
 p(F_2 |\map B \beta)
 &= p(F_2 | \map B \map A'_1 \overline{E}_1\map A_1 \beta ) \,
 p( \overline{E}_1 |\map A_1\beta)\\
 &= p(E_2| \map A_2 \map A'_1 \overline{E}_1\map A_1 \beta )\,
 p( \overline{E}_1 |\map A_1\beta)
\end{split}\end{equation}
for every belief $\beta$ with $p(\overline{E}_1|\map A_1\beta)\ne 0$. Since $p(F_2|\map B\beta)= p(E_2|\map A_2\beta)$ this yields Eq.~\eqref{lem5}.

If $p(\overline{E}_1|\map A_1\beta)=0$, then Eq.~\eqref{emab2} yields 
immediately $p(E_2|\map A_2\beta)=p(F_2|\map B\beta)= 0$.
\qed

\medskip 

We are now ready to prove the following formulation of Theorem~\arabic{thm2}:

\begin{theobis}
Let $\set{(\map A_n, E_n)| n =1,\dotsc,k}$ be a set of ideal outcomes which are 
pairwise mutually exclusive. Then the condition
\begin{equation}\label{exclu}
 \sum_{n=1}^k p(E_n|\map A_n\beta)\le 1
\end{equation}
is satisfied for every belief $\beta$.
\end{theobis}

{\bf Proof.}  The idea of the proof is to construct a sequential experiment with the property that a subset of outcomes of this experiment have the same probabilities of the ideal outcomes $\set{(\map A_n, E_n)| n =1,\dotsc,k}$. Since the sum of the probabilities over all outcomes of the sequential measurement is 1, this will imply Eq.~(\ref{exclu}).

We define $\Part E_i=(E_i,\overline{E}_i)$, and denote by $\map A'_i$ denote the reversing action of $A_i$, as in Eq.~(4) of the main text. Consider the sequential experiment defined by the following 
procedure:
\begin{itemize}
\item[(I)]
Move to step ($R_1$).
\item[($R_i$)]
Perform the action $(\map A_i, \Part E_i, \map A'_i)$.
If the event  $E_i$ takes place, then move to step ($T_i$).
If $i = k$ and the event $\overline{E}_i$ takes place, then move to step 
($T_{k+1}$).
Otherwise, move to step ($R_{i+1}$).
\item[($T_s$)]
Report outcome $s$ and terminate.
\end{itemize}
The above procedure defines a sequential experiment with outcome $s  \in \{1,2,\dotsc, k+1\}$. 
We denote by $q_\beta(s)$ the probability assigned to the outcome $s$ when the initial belief is $\beta$.

We now prove the equality 
\begin{align}\label{desired}q_\beta(s)= p(E_s|\map A_s\beta)  \qquad \forall s \in  \{1,
\dots, k\} \, .
\end{align} 
For $s=1$, the equality (\ref{desired}) follows by the definition of the sequential experiment, which has  $q_\beta (1) =  p(E_1|\map A_1\beta)$.

To deal with the $s>1$ case,  we  define the sequence of beliefs $\beta_0,\beta_1,\dotsc,\beta_k$, via the relations   
$\beta_0 :=\beta$ and
\begin{equation}
\beta_j :=\map A'_j\, \overline{E}_j\, \map A_j\beta_{j-1}  \qquad \forall j\in  \{1,\dots, k\} 
\end{equation}
With this definition, the belief $\beta_{i-1}$ is the belief of the agent at the beginning of 
step ($R_i$).    For the purpose of this proof, if for some $j$ the probability $p( \overline{E}_j | \map A_j  \beta_{j-1})$ is zero and  the updated belief $\map A'_j\, \overline{E}_j\, \map A_j\beta_{j-1}$,  then $\beta_j$ can be chosen to be an arbitrary belief.

The probability of an outcome $1<s\le k$ is 
\begin{align}
q_\beta(s)&=p(E_s|\map A_s\beta_{s-1})\prod_{j=1}^{s-1}  p(\overline E_j|\map A_j\beta_{j-1}) \, .\label{qs}
\end{align}

To prove Eq. (\ref{desired}) in  the case $s\ge 2$,  we apply Lemma~\ref{lem:sameeffect} to the mutually exclusive ideal outcomes $(\map A_s,E_s)$ and $(\map A_j, 
E_j)$ for $j\in  \{1,\dots, s-1\}$. 
The key observation is  that Lemma~\ref{lem:sameeffect}  implies the equality  
\begin{align}
p(E_s |\map A_s  \, \beta_j  )  \,  p(\overline{E}_{j} |\map A_{j}\beta_{j-1}) &= p(\overline{E}_{s} |\map A_{s}\beta_{j-1})   \label{equa} 
\end{align}
valid for every $ j\in \{1,\dots, s-1\}$. For $p( \overline{ E}_j |  \map A_j \beta_{j-1}) \not =0$, Eq.~(\ref{equa}) follows from the definition $\beta_j:  =  \map A_{j}'  \overline E_j  \map A_j  \beta_{j-1}$, and from the first implication in Lemma~\ref{lem:sameeffect}.  For $p( \overline{ E}_j |  \map A_j \beta_{j-1}) \not =0$, Eq.~(\ref{equa}) follows from the second implication in Lemma~\ref{lem:sameeffect}.

By using Eq.~(\ref{equa}),   we obtain
\begin{align}
 \nonumber q_\beta(s)&=p(E_s|\map A_s\beta_{s-1})\prod_{j=1}^{s-1} \,  p(\overline E_j|\map A_j\beta_{j-1})\\
\nonumber  &= p(E_s |\map A_s  \beta_{s-1} ) \,  p(\overline{E}_{s-1} |\map A_{s-1}\beta_{s-2})\\
\nonumber & \qquad \times \prod_{j=1}^{s-2}(1-q_\beta(j))\\
\nonumber &= p(E_s |\map A_s\beta_{s-2}) \prod_{j=1}^{s-2}   p(\overline E_j|\map A_j\beta_{j-1}) \\
\nonumber &\quad\vdots\\
 \nonumber &= p(E_s |\map A_s\beta_0) \\
 &= p(E_s |\map A_s\beta)\,  ,  \label{nzero}
 \end{align}
where the third equality follows from Eq.~(\ref{equa}) with $j=s-1$, and the subsequent iterations follow from  Eq.~(\ref{equa})  with $j$ starting from  $s-2$ and going down to $j=1$. 

Hence, we have proven Eq.~(\ref{desired}).  Using Eq.~(\ref{desired}), it is now easy to prove Eq.~(\ref{exclu}), and therefore the validity of the exclusivity principle.  Since the probabilities  $(q_\beta (j) \,)_{j=1}^{k+1}$ sum up to 1,  one simply has 
\begin{equation}
\sum_{s=1}^k p(E_s|\map A_s\beta)=\sum_{s=1}^k q_\beta(s)\le 1\,.
\end{equation}
\qed



\begin{thebibliography}{99}



\bibitem{Everett57}
H. Everett, III,
``Relative state'' formulation of quantum mechanics,
\href{https://doi.org/10.1103/RevModPhys.29.454}{Rev. Mod. Phys. \textbf{29}, 454 (1957).}


\bibitem{FS13}
C. A. Fuchs and R. Schack,
Quantum-Bayesian coherence,
\href{https://doi.org/10.1103/RevModPhys.85.1693}{Rev. Mod. Phys. \textbf{85}, 1693 (2013).}

\bibitem{bell1964einstein}
J. S. Bell,
On the Einstein Podolsky Rosen paradox,
\href{https://doi.org/10.1103/PhysicsPhysiqueFizika.1.195}{Physics {\bf 1}, 195 (1964).}

\bibitem{khalfin1992quantum}
L. A. Khalfin and B. S. Tsirelson,
Quantum/classical correspondence in the light of Bell's inequalities,
\href{https://doi.org/10.1007/BF01889686}{Found. Phys. \textbf{22}, 879 (1992).}

\bibitem{popescu1994quantum}
S. Popescu and D. Rohrlich,
Quantum nonlocality as an axiom,
\href{https://doi.org/10.1007/BF02058098}{Found. Phys. {\bf 24}, 379 (1994).}

\bibitem{rastall1985locality}
P. Rastall,
Locality, Bell's theorem, and quantum mechanics,
\href{https://doi.org/10.1007/BF00739036}{Found. Phys. {\bf 15}, 963 (1985).}

\bibitem{brassard2006limit}
G. Brassard, H. Buhrman, N. Linden, A. A. M\'ethot, A. Tapp, and F. Unger,
Limit on nonlocality in any world in which communication complexity is not trivial,
\href{https://doi.org/10.1103/PhysRevLett.96.250401}{Phys. Rev. Lett. {\bf 96}, 250401 (2006).}

\bibitem{linden2007noquantum}
N. Linden, S. Popescu, A. J. Short, and A. Winter,
Quantum nonlocality and beyond: Limits from nonlocal computation,
\href{https://doi.org/10.1103/PhysRevLett.99.180502}{Phys. Rev. Lett. {\bf 99}, 180502 (2007).}

\bibitem{pawlowski2009information}
M. Paw{\l}owski, T. Paterek, D. Kaszlikowski, V. Scarani, A. Winter, and M. \.Zukowski,
Information causality as a physical principle,
\href{https://doi.org/10.1038/nature08400}{Nature {\bf 461}, 1101 (2009).}

\bibitem{navascues2010glance}
M. Navascu\'es and H. Wunderlich,
A glance beyond the quantum model,
\href{https://doi.org/10.1098/rspa.2009.0453}{Proc. R. Soc. A \textbf{466}, 881 (2009).}

\bibitem{fritz2013local}
T. Fritz, A. B. Sainz, R. Augusiak, J. Bohr Brask, R. Chaves, A. Leverrier, and A. Ac\'in,
Local orthogonality as a multipartite principle for quantum correlations,
\href{https://doi.org/10.1038/ncomms3263}{Nat. Commun. {\bf 4}, 2263 (2013).}

\bibitem{CSW10}
A. Cabello, S. Severini, and A. Winter,
(Non\mbox{-})con\-tex\-tu\-al\-i\-ty of physical theories as an axiom,
\href{https://arxiv.org/abs/1010.2163}{\eprint{arXiv:1010.2163}.}

\bibitem{Cabello13}
A. Cabello,
Simple explanation of the quantum violation of a fundamental inequality,
\href{https://doi.org/10.1103/PhysRevLett.110.060402}{Phys. Rev. Lett. \textbf{110}, 060402 (2013).}

\bibitem{Yan13}
B. Yan,
Quantum correlations are tightly bound by the exclusivity principle,
\href{https://doi.org/10.1103/PhysRevLett.110.260406}{Phys. Rev. Lett. \textbf{110}, 260406 (2013).}

\bibitem{ATC14}
B. Amaral, M. Terra Cunha, and A. Cabello,
Exclusivity principle forbids sets of correlations larger than the quantum set,
\href{https://doi.org/10.1103/PhysRevA.89.030101}{Phys. Rev. A \textbf{89}, 030101(R) (2014).}

\bibitem{Henson15}
J. Henson,
Bounding quantum contextuality with lack of third-order interference,
\href{https://doi.org/10.1103/PhysRevLett.114.220403}{Phys. Rev. Lett. \textbf{114}, 220403 (2015).}

\bibitem{AFLS15}
A. Ac\'{\i}n, T. Fritz, A. Leverrier, and A. B. Sainz,
A combinatorial approach to nonlocality and contextuality,
\href{https://doi.org/10.1007/s00220-014-2260-1}{Comm. Math. Phys. \textbf{334}, 533 (2015).}

\bibitem{Cabello15}
A. Cabello,
Simple explanation of the quantum limits of genuine $n$-body nonlocality,
\href{https://doi.org/10.1103/PhysRevLett.114.220402}{Phys. Rev. Lett. \textbf{114}, 220402 (2015).}

\bibitem{Cabello18}
A. Cabello, Quantum correlations from simple assumptions, \href{https://journals.aps.org/pra/abstract/10.1103/PhysRevA.100.032120}{Phys. Rev. A \textbf{100}, 032120 (2019).}

\bibitem{CY16}
G. Chiribella and X. Yuan,
Bridging the gap between general probabilistic theories and the device-independent framework for nonlocality and contextuality,
\href{https://doi.org/10.1016/j.ic.2016.02.006}{Inf. Comput. {\bf 250}, 15 (2016).}





\bibitem{kraemer2018operational}
L. Kr\"amer and L. del Rio,
Operational locality in global theories,
\href{https://doi.org/10.1098/rsta.2017.0321}{Phil. Trans. R. Soc. A {\bf 376}, 20170321 (2018).}

\bibitem{chiribella2018agents}
G. Chiribella,
Agents, subsystems, and the conservation of information,
\href{https://doi.org/10.3390/e20050358}{Entropy {\bf 20}, 358 (2018).}



\bibitem{CY14}
G. Chiribella and X. Yuan,
Measurement sharpness cuts nonlocality and contextuality in every physical theory,
\href{https://arxiv.org/abs/1404.3348}{\eprint{arXiv:1404.3348}.}


\bibitem{MK14}
M. Kleinmann,
Sequences of projective measurements in generalized probabilistic models,
\href{https://doi.org/10.1088/1751-8113/47/45/455304}{J. Phys. A \textbf{47}, 455304 (2014).}

\bibitem{Fuchs}
C. A. Fuchs,
Quantum mechanics as quantum information, mostly,
\href{https://doi.org/10.1080/09500340308234548}{J. Mod. Opt. \textbf{50}, 987 (2003).}

\bibitem{Hardy01}
L. Hardy,
Quantum theory from five reasonable axioms,
\href{https://arxiv.org/abs/quant-ph/0101012}{\eprint{quant-ph/0101012}.}

\bibitem{CDP11}
G. Chiribella, G. M. D'Ariano, and P. Perinotti,
Informational derivation of quantum theory,
\href{https://doi.org/10.1103/PhysRevA.84.012311}{Phys. Rev. A \textbf{84}, 012311 (2011).}

\bibitem{DB11}
B. Daki\'c and \v{C}. Brukner,
Quantum theory and beyond: Is entanglement special?,
in \emph{Deep Beauty. Understanding the Quantum World through Mathematical Innovation},
edited by H. Halvorson
(Cambridge University Press, New York, 2011), p.~365.

\bibitem{MM11}
L. Masanes and M. P. M\"uller,
A derivation of quantum theory from physical requirements,
\href{https://doi.org/10.1088/1367-2630/13/6/063001}{New J. Phys. \textbf{13}, 063001 (2011).}

\bibitem{Hardy11}
L. Hardy,
Reformulating and reconstructing quantum theory,
\href{https://arxiv.org/abs/1104.2066}{\eprint{arXiv:1104.2066}.}

\bibitem{Wilce}
A. Wilce,
A royal road to quantum theory (or thereabouts),
\href{https://arxiv.org/abs/1606.09306}{\eprint{arXiv:1606.09306}.}

\bibitem{HW}
P. A. H\"ohn and C. S. P. Wever,
Quantum theory from questions,
\href{https://doi.org/10.1103/PhysRevA.95.012102}{Phys. Rev. A \textbf{95}, 012102 (2017).}


\end{thebibliography}
\end{document}